\documentclass[a4paper,12pt]{article}

\usepackage[utf8x]{inputenc}
\usepackage[T1]{fontenc}

\usepackage[colorlinks=true,linkcolor=black,citecolor=black,urlcolor=blue]{hyperref}
\usepackage{geometry}
\usepackage{titlesec}
\titleformat{\section}
{\bfseries\uppercase}{\thesection.}{1em}{}
\titleformat{\subsection}
{\bfseries}{\thesection.\thesubsection.}{1em}{}

\usepackage{graphicx} 
\usepackage{multirow} 
\usepackage{cite}
\usepackage{breakurl}
\usepackage{indentfirst}
\usepackage{amsmath, amssymb, amsfonts, bm}
\usepackage{txfonts}
\usepackage{fourier}
\usepackage{enumitem}
\usepackage{xcolor}
\usepackage{enumitem}
\usepackage{tabularx} 
\usepackage{makecell} 

\titlespacing*{\subsection}{0pt}{1.5em}{0.2em}
\titlespacing*{\section}{0pt}{1.5em}{0.2em}

\renewcommand\eqref[1]{Equation~\ref{#1}}

\renewcommand{\thesection}{\arabic{section}}
\renewcommand{\thesubsection}{\arabic{subsection}}

\makeatletter
\renewcommand\@biblabel[1]{#1.}
\makeatother

\newlength{\bibitemsep}
\newlength{\bibparskip}
\let\oldthebibliography\thebibliography
\renewcommand\thebibliography[1]{%
  \oldthebibliography{#1}%
  \setlength{\parskip}{\bibitemsep}%
  \setlength{\itemsep}{\bibparskip}%
}

\newcommand{\YearConf}{2024}

\newcommand{\LogoConf}{logo_IN24.jpg}

\usepackage{fancyhdr}
  \fancypagestyle{firststyle}
{   \fancyhf{}
}
\begin{document}
\thispagestyle{firststyle}

\begin{center}
	\includegraphics[width=2in]{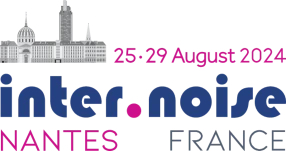}
\end{center}
\vskip.5cm

\begin{flushleft}
\fontsize{16}{20}\selectfont\bfseries

\color{black}Computation-efficient Virtual Sensing Approach with Multichannel Adjoint Least Mean Square Algorithm
\end{flushleft}
\vskip1cm
\renewcommand\baselinestretch{1}
\begin{flushleft}

Boxiang Wang\footnote{boxiang001@e.ntu.edu.sg}, Junwei Ji\footnote{junwei002@e.ntu.edu.sg}, Xiaoyi Shen\footnote{xiaoyi.shen@ntu.edu.sg}, Dongyuan Shi\footnote{dongyuan.shi@ntu.edu.sg}, Woon-Seng Gan\footnote{ewsgan@ntu.edu.sg}\\
Digital Signal Processing Lab, Nanyang Technological University\\
50 Nanyang Avenue, Singapore 639798\\

\end{flushleft}
\textbf{\centerline{ABSTRACT}}\\
\textit{Multichannel active noise control (ANC) systems are designed to create a large zone of quietness (ZoQ) around the error microphones, however, the placement of these microphones often presents challenges due to physical limitations. Virtual sensing technique that effectively suppresses the noise far from the physical error microphones is one of the most promising solutions. Nevertheless, the conventional multichannel virtual sensing ANC (MVANC) system based on the multichannel filtered reference least mean square (MCFxLMS) algorithm often suffers from high computational complexity. This paper proposes a feedforward MVANC system that incorporates the multichannel adjoint least mean square (MCALMS) algorithm to overcome these limitations effectively. Computational analysis demonstrates the improvement of computational efficiency and numerical simulations exhibit comparable noise reduction performance at virtual locations compared to the conventional MCFxLMS algorithm. Additionally, the effects of varied tuning noises on system performance are also investigated, providing insightful findings on optimizing MVANC systems.}

\section{INTRODUCTION}
\noindent
Active noise control (ANC) is an advanced technique that can effectively suppress the unwanted sound by the anti-noise, which has the same amplitude and opposite phase of the original noise \cite{nelson1991active,elliott1993active,hansen1999understanding,qiu2019introduction}. To react to the noise and environment changes, the filtered reference least mean square (FxLMS) algorithm is one of the most practical algorithms that is widely used in common adaptive ANC systems \cite{morgan1980analysis,kuo1996active,kajikawa2012recent}. Its extension has been proposed in these years to address some practical problems \cite{shi2024behind,ji2023practical,luo2023delayless,lai2023mov,okajima2022dual,sun2020realistic}. Due to its efficiency in reducing low-frequency noise and compact size, ANC is widely utilized in many applications, such as headphones~\cite{shen2021wireless,shen2022multi,shen2022hybrid,shen2022adaptive}, headrests~\cite{rafaely1999broadband,chang2022multi,zhang2022robust,shen2023implementations}, and windows~\cite{shi2016open,shi2017algorithms,lam2018active,lam2020active,lam2020active1,lam2023anti,luo2024real}.

The placement of error microphones is critical to the effectiveness of an ANC system. Generally, the error microphones are placed in the desired zone of quietness (ZoQ) to monitor the residual error signal, leading to a gradual reduction in the combined acoustic power of the anti-noise and the undesirable noise to its lowest level. For example, within an ANC headrest, it is ideal for the ZoQ to encompass the area around the user's eardrum. However, positioning the error microphone near the user’s eardrum to capture signals is impractical, preventing the formation of a quiet zone around the ears. Consequently, in situations where it is not feasible to position error microphones within the ZoQ, obtaining precise estimations of error signals becomes indispensable for maintaining the ANC system's performance. A promising solution to this challenge is the virtual sensing technique \cite{toyooka2022online,toyooka2023hybrid,moreau2008review,zhang2021robust}, which effectively generates the quiet zone at desired ‘virtual’ locations far from the physical error sensor. To date, a variety of virtual sensing ANC techniques have been developed, which broadly fall into two main categories~\cite{liu2009virtual}. Class I techniques employ direct estimation of sound characteristics at the virtual location using the physical error microphone, based on either acoustical models~\cite{garcia1997generation} or extrapolating polynomials~\cite{kestell2001active}. However, the efficacy of these approaches in attenuating noise is considerably dependent on the accuracy of the models used~\cite{garcia1997generation}, limiting their applicability primarily to low-frequency tonal sound environments~\cite{moreau2009active}. Class II virtual sensing techniques necessitate a preliminary training phase to effectively manage noise. Within this category, the remote microphone technique computes an 'observation filter' that maps the sound from the physical microphone to the targeted location~\cite{jung2017local,jung2019local}. This method, however, demands careful placement of the physical error microphones. Alternatively, the auxiliary-filter-based approach implicitly captures the relationship between physical and virtual microphones, thus avoiding the causality constraint~\cite{hasegawa2018window,shi2019analysis,shi2020feedforward}.

Extensive research efforts have been invested in applying the auxiliary-filter-based virtual sensing technique across a wide range of ANC applications. Pawelczyk introduced this technique for ANC headrests through both single and multichannel configurations, employing adaptive feedback systems~\cite{pawelczyk2004adaptive}. The efficacy of Pawelczyk’s system was further validated by Hasegawa et al. in tonal noise control~\cite{hasegawa2017headrest}, and by Hirose et al. in factory noise control~\cite{hirose2017effectiveness} using ANC headrests, while Miyazaki et al. implemented the virtual sensing technique on a head-mounted ANC system~\cite{miyazaki2015head}. Additionally, Deng et al. expanded upon this work by developing a $1 \times 2 \times 2$ feedforward structure using the virtual sensing technique for ANC headrests~\cite{deng2018active}.  
These studies demonstrate that a multichannel virtual sensing ANC (MVANC) structure is crucial for establishing a broad ZoQ and enhancing the system's adaptability to complex acoustic environments, especially in open spaces. Nevertheless, employing the traditional FxLMS algorithm for adaptive control filter updates significantly escalates computational complexity as the number of channels rises~\cite{kuo1996active}. Additionally, the virtual sensing technique incurs extra computational demands during the auxiliary filter’s training and execution phases~\cite{shi2020feedforward}. These shortcomings pose significant challenges to the feasibility of MVANC systems in real-world ANC applications.

This paper explores a feedforward MVANC system that employs the multichannel adjoint least mean square (MCALMS) algorithm~\cite{wan1996adjoint,shi2023computation} instead of the standard multichannel filtered-x least mean square (MCFxLMS) algorithm to adaptively update the control filters, which enhances the computation efficiency significantly while maintaining the same level of noise reduction as the MCFxLMS algorithm. This paper is organized as follows: Section $2$ outlines the development of the feedforward MVANC system with the MCALMS algorithm. Section $3$ provides a comprehensive computational analysis of the proposed algorithm. Simulations of a $4 \times 2 \times 4$ feedforward MVANC system are conducted in Section $4$ to evaluate the algorithm's effectiveness and the impacts of various tuning noises. Finally, a conclusion is given in Section $5$.

\section{Feedforward MVANC system with MCALMS algorithm}
\noindent
In this section, the feedforward MVANC system employing the MCALMS algorithm is derived to work with error microphones placed far from the desired zone of quietness. The virtual sensing technique includes two stages: the tuning stage and the control stage. In the tuning stage, the

\begin{figure}[h!]
\begin{center}
  \includegraphics[width=4in]{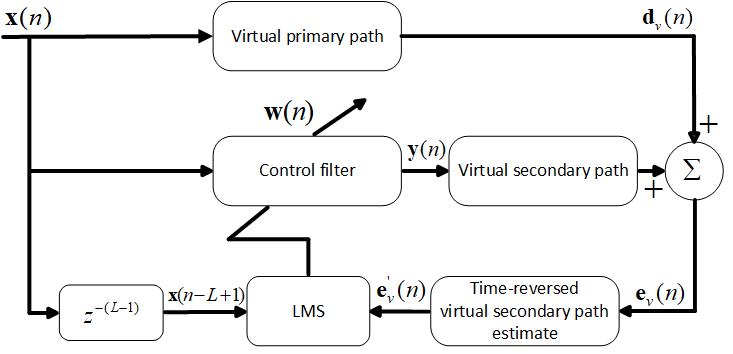}
  \end{center}
  \caption{Tuning stage: block diagram of training the optimal control filters.}
  \label{fig:1}
\end{figure}

\noindent
auxiliary filters are trained to record the information between physical microphones and virtual microphones indirectly. These trained auxiliary filters then play a crucial role in the control stage, facilitating noise suppression at the designated virtual locations. Specifically, the MCALMS algorithm is combined with the virtual sensing technique to update the control filters, which reduces the system’s computation complexity. In this paper, we focused on a feedforward MVANC system with $J$ reference microphones, $K$ secondary sources, $M$ physical microphones and $Q$ virtual microphones.

\subsection{The tuning stage}
\noindent
The auxiliary filters, which implicitly make up the difference between physical and virtual microphones, are crucial to the MVANC system. Before estimating it, the optimal control filters for the virtual error microphones at the desired noise reduction area should be first obtained as shown in Fig.~\ref{fig:1}. The $k$th control signal is given by:
\begin{equation}
{y}_k(n) = \sum_{j=1}^J\mathbf{w}_{kj}^\text{T}(n)\mathbf{x}_{j}(n), \; k = 1,2,\cdots,K,
\label{Eq:1}
\end{equation}
where $\mathbf{w}_{kj}(n) = [w_{kj,0}(n)\, w_{kj,1}(n)\, \ldots\, w_{kj,N_x-1}(n)]^\text{T}$ is the control filter from the $j$th input to the $k$th output with the length of $N_x$ and $\mathbf{x}_j(n) = [x_j(n)\, x_j(n - 1)\, \ldots\, x_j(n - N_x + 1)]^\text{T}$ denotes the $j$th reference vector at time index $n$. Then, these control signals will be played by the secondary sources to suppress the disturbance at virtual microphones, resulting in:
\begin{equation}
e_{v,q}(n) = d_{v,q}(n) - \sum_{k=1}^K y_k(n)*{s}_{v,qk}, \; q = 1,2,\cdots,Q,
\label{Eq:2}
\end{equation}
where $d_{v,q}(n)$ is the disturbance signal captured by the $q$th virtual error sensor and $s_{v,qk}$ denotes the virtual secondary path from the $k$th secondary source to the $q$th virtual error sensor.

The control filters keep updating their coefficients until the error signal is minimized to obtain the optimum value. However, the conventional MCFxLMS places a high cost on the computation. To improve the efficiency of the system, the MCALMS algorithm \cite{wan1996adjoint,shi2023computation} is introduced to reduce the computational burden. In contrast to the MCFxLMS algorithm where the reference signals are filtered by the estimated secondary paths, the MCALMS algorithm filters the error signal instead. Assuming that the length of the estimated secondary path is $L$. To ensure causality, the past $L-1$ samples should be taken into account. As a result, the control filter is updated as:
\begin{equation}
\mathbf{w}_{kj}(n + 1) = \mathbf{w}_{kj}(n) - \mu_1 \mathbf{x}_j(n - L + 1) \sum_{q=1}^{Q} e'_{v,kq}(n),
\label{Eq:3}
\end{equation}
\begin{figure}[h!]
\begin{center}
  \includegraphics[width=4in]{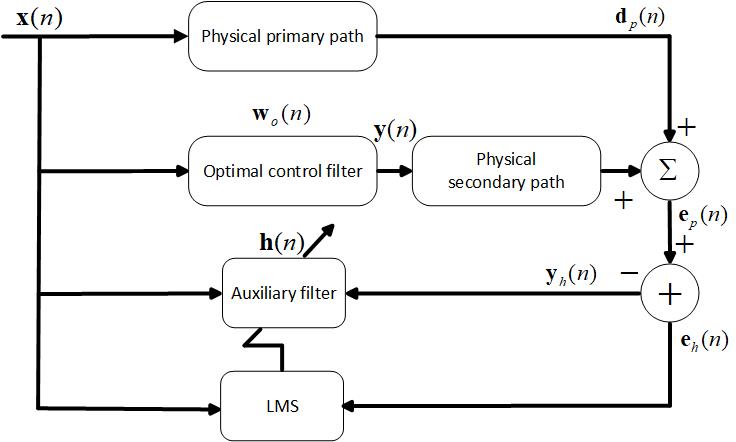}
  \end{center}
  \caption{Tuning stage: Block diagram of training the optimal auxiliary filters.}
  \label{fig:2}
\end{figure}

\noindent
where $\mu_1$ denotes the stepsize of the MCALMS algorithm and $e'_{v,kq}(n)$ represents the time-reversed filtered error signal, which is obtained by:
\begin{equation}
e'_{v,kq}(n) = \sum_{i=0}^{L-1} e_{v,q}(n - L + 1 + i)\hat{s}_{v,qk,i},
\label{Eq:4}
\end{equation}
where $\hat{s}_{v,qk,i}$ represents the $i$th coefficient of the impulse response of the estimated virtual secondary path from the $k$th secondary source to the $q$th virtual error sensor.

Once the optimal control filters are obtained, the auxiliary filters will be trained using the least mean square (LMS) algorithm as shown in Fig.~\ref{fig:2}. In this case, the virtual microphones are replaced by the physical microphones. Hence, the auxiliary filter vector is updated by:
\begin{equation}
\mathbf{h}_{mj}(n + 1) = \mathbf{h}_{mj}(n) + \mu_2 e_{h,m}(n) \mathbf{{\bar x}}_j(n),
\label{Eq:5}
\end{equation}
where $\mu_2$ denotes the stepsize of the LMS algorithm. $\mathbf{h}_{mj}(n)$ represents the auxiliary filter from the $j$th reference sensor to the $m$th physical error sensor of tap length $N_h$ and the $j$th reference vector $\mathbf{\bar x}_j(n)$ is expressed as:
\begin{equation}
\bar{\mathbf{x}}_j(n) = \left[
x_j(n)\, \,
x_j(n - 1)\, \,
\ldots\, \,
x_j(n - N_h + 1)
\right]^\text{T}.
\label{Eq:6}
\end{equation}

The inner error signal $e_{h,m}(n)$ shown in \eqref{Eq:5} is stated as:
\begin{equation}
    e_{h,m}(n) = e_{p,m} - \sum_{j=1}^J\mathbf{h}_{mj}^\text{T}(n)\mathbf{\bar{x}}_j(n),
    \label{Eq:7}
\end{equation}
where $e_{p,m}$ represents the residual signal captured at the $m$th physical error sensor.

In the tuning stage, the optimal control filters for the virtual error microphone are obtained first, and subsequently, these optimal controllers will be used to estimate the auxiliary filter as a way of compensating for the difference in sound field between the virtual error microphones and the physical error microphones. The auxiliary filters will then help the MVANC system to provide effective noise cancellation of various noises at the virtual error microphone positions during the control stage.

\begin{figure}[h!]
\begin{center}
  \includegraphics[width=4in]{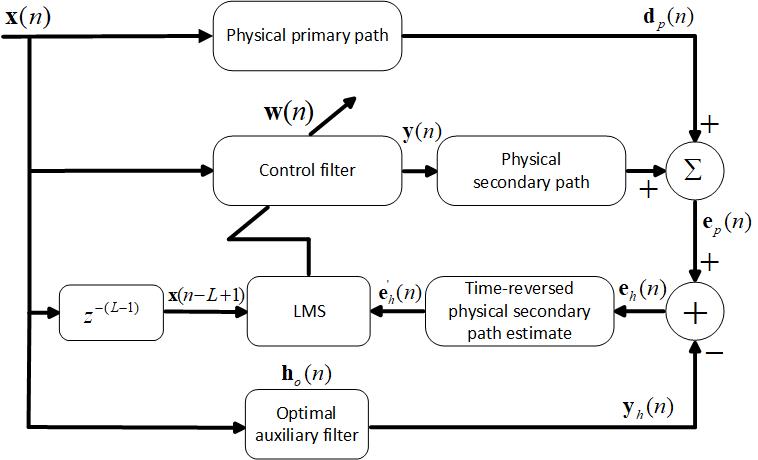}
  \end{center}
  \caption{Control stage: Block diagram of training the control filters.}
  \label{fig:3}
\end{figure}

\subsection{The control stage}
\noindent
During the control stage, the physical microphones remain while the virtual microphones are removed. The pre-trained optimal auxiliary filters are introduced into the conventional MCALMS algorithm as shown in Fig.~\ref{fig:3}. Therefore, the update equation for the control filter at the control stage becomes:
\begin{equation}
\mathbf{w}_{kj}(n + 1) = \mathbf{w}_{kj}(n) - \mu_3 \mathbf{x}_j(n - L + 1) \sum_{m=1}^{M} e'_{h,km}(n),
\label{Eq:8}
\end{equation}
where $\mu_3$ is the stepsize of the MCALMS algorithm and $e'_{h,km}(n)$ represents the time-reversed filtered inner error signal, which is obtained by:
\begin{equation}
e'_{h,km}(n) = \sum_{i=0}^{L-1} e_{h,m}(n - L + 1 + i)\hat{s}_{p,mk,i},
\label{Eq:9}
\end{equation}
where $\hat{s}_{p,mk,i}$ represents the $i$th coefficient of the impulse response of the estimated physical secondary path between the $k$th secondary source and $m$th physical error microphone. The inner error signal $e_{h,km}$ is deduced as:
\begin{equation}
e_{h,m}(n) = e_{p,m}(n) - \sum_{j=1}^J\mathbf{h}_{o,mj}^T \bar{\mathbf{x}}_j(n),
\label{Eq:10}
\end{equation}
where $\mathbf{h}_{o,mj}$ denotes the optimal auxiliary filter from the $j$th input to the $m$th physical microphone.

The MVANC system leverages auxiliary filters to create noise reduction areas where physical error microphones are difficult to place. MCALMS dramatically reduces the computational effort and storage space compared to the traditional algorithms, which in turn facilitates real-time implementations. Following the tuning stage and control stage stated above, the proposed feedforward MVANC system can realize noise cancellation at $Q$ virtual positions.





\section{Computational Analysis of MVANC system with MALMS algorithm}
\noindent
This section examines the computational complexity of the MVANC system employing the MCALMS algorithm, contrasting it with the computational demands of the MVANC system that uses the MCFxLMS algorithm. Considering the tuning phase is conducted offline, this analysis focuses solely on the computational demands during the control stage. Each algorithm requires five operations at this stage. Utilizing these operations, we quantify the necessary additions and multiplications for both algorithms within a single cycle of the control stage, as detailed in Table~\ref{Tab:1}. It can be seen from Table~\ref{Tab:1} that, compared to the conventional MCFxLMS, the MCALMS algorithm requires less multiplication and addition. This reduction is attributed to the efficiency of the filtered error mechanism in updating the control filters.

\begin{table}[h]
\caption{The number of multiplications and additions used in the MVANC system with MCFxLMS and MCALMS algorithms for one period.}
\centering
\begin{tabularx}{\textwidth}{|X|X|X|}
\hline
\textbf{Algorithm} & \textbf{Multiplications} & \textbf{Additions} \\
\hline
MVANC + MCFxLMS & \(JKM(L + N_x + 1) + MJN_h\) & \(JKM(L + N_x - 1) + M(J + N_h - 1)\) \\
\hline
MVANC + MCALMS & \(K(LM + JN_x + 1) + MJN_h\) & \(K[(L - 1)M + J(N_x + M - 1)] + M(J + N_h - 1)\) \\
\hline
\end{tabularx}
\label{Tab:1}
\end{table}

\begin{figure}[h!]
\begin{center}
  \includegraphics[width=5in]{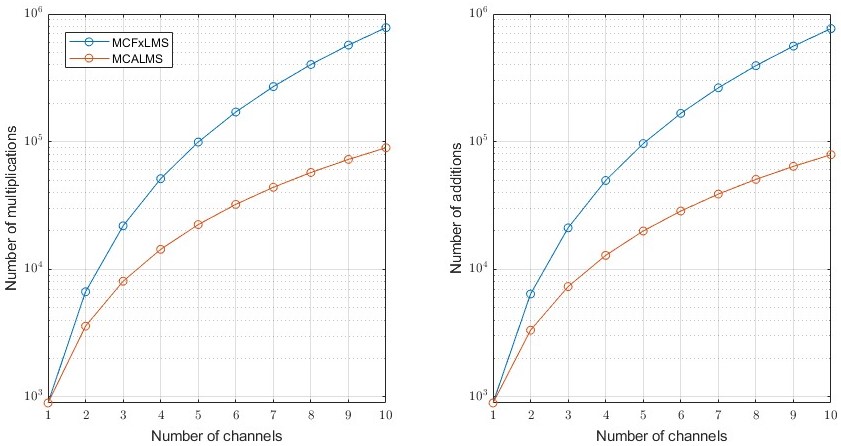}
  \end{center}
  \caption{Multiplications and additions of the MVANC system with MCFxLMS and MCALMS algorithms.}
  \label{fig:4}
\end{figure}
\noindent

To underscore the computational efficiency of the MVANC system using the MCALMS algorithm, parameters $N_x$, $N_h$, and $L$ are set to $512$, $128$, and $256$ respectively, with $J$, $K$, and $M$ representing the number of channels, which varies from $1$ to $10$. The computational load of both algorithms, in relation to the increasing channel count, is graphically illustrated in Fig.~\ref{fig:4}. This visual representation clearly demonstrates that the MCALMS algorithm achieves considerable computational savings compared to the MCFxLMS algorithm as the number of channels increases. Specifically, when the number of channels reaches $10$, the computational load of the MCFxLMS algorithm is almost 10 times that of the MCALMS algorithm.

\begin{figure}[h!]
\begin{center}
  \includegraphics[width=4in]{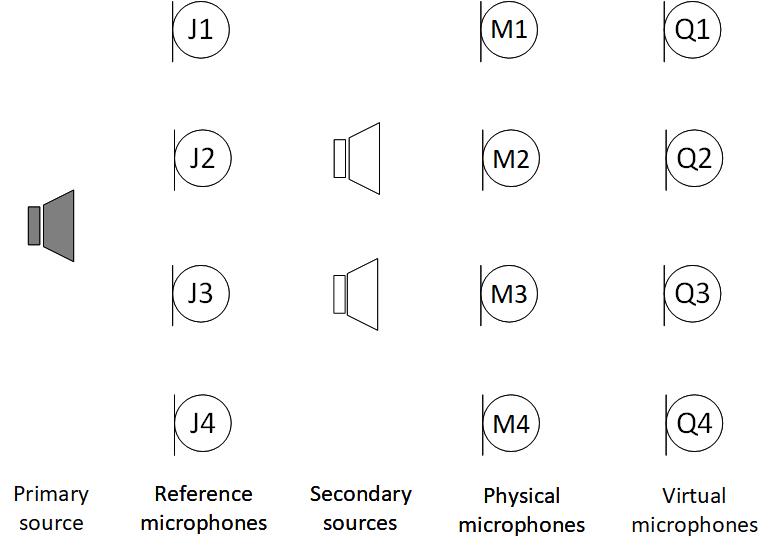}
  \end{center}
  \caption{Schematic diagram of the $4 \times 2 \times 4$ MVANC system.}
  \label{fig:5}
\end{figure}

\section{Simulation results}
\noindent
To verify the effectiveness of the proposed MVANC system utilizing the MCALMS algorithm, as well as to explore the impact of different tuning noises on the noise reduction performance in the control stage. Two simulations are conducted on a fully connected ANC system, where there are four reference microphones ($J = 4$), two secondary sources ($K = 2$), four physical microphones ($M = 4$) and four virtual microphones ($Q = 4$), as shown in Fig.~\ref{fig:5}. The physical and virtual primary and secondary paths are bandpass filters with a frequency range of $500$ to $5000$ Hz. The control filters, auxiliary filters, primary path, and secondary path models are configured with $512$, $256$, $128$, and $32$ taps, respectively. The system's sampling rate is established at $16$ kHz. To mimic a realistic environment, white noises are introduced to the primary noises detected by the reference microphones with a signal-to-noise ratio of 40 dB.
\begin{figure}[h!]
\begin{center}
  \includegraphics[width=5in]{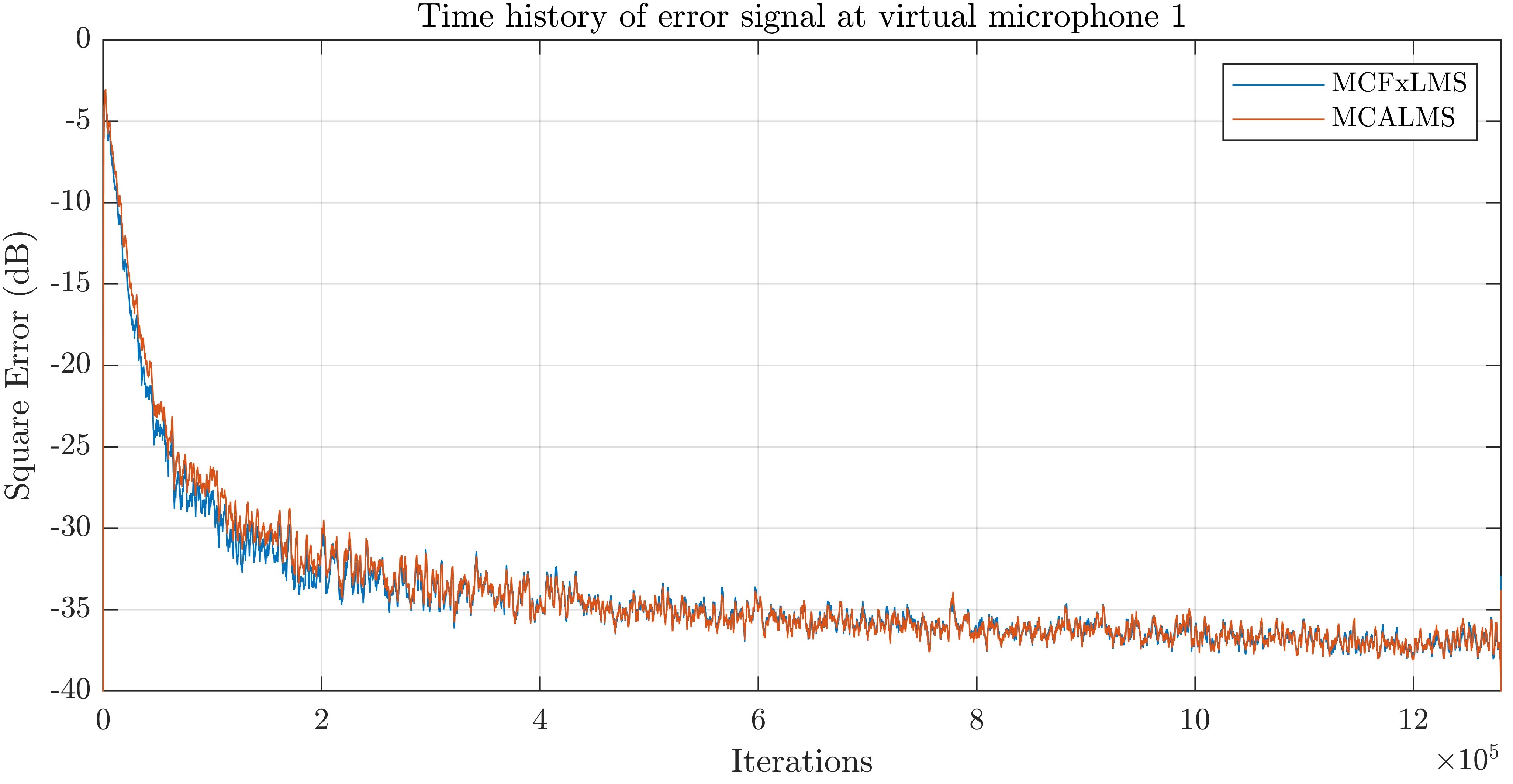}
  \end{center}
  \caption{Error signal at virtual microphone 1 of the MVANC system with MCFxLMS and MCALMS algorithms.}
  \label{fig:6}
\end{figure}

\begin{figure}[h!]
\begin{center}
  \includegraphics[width=4in]{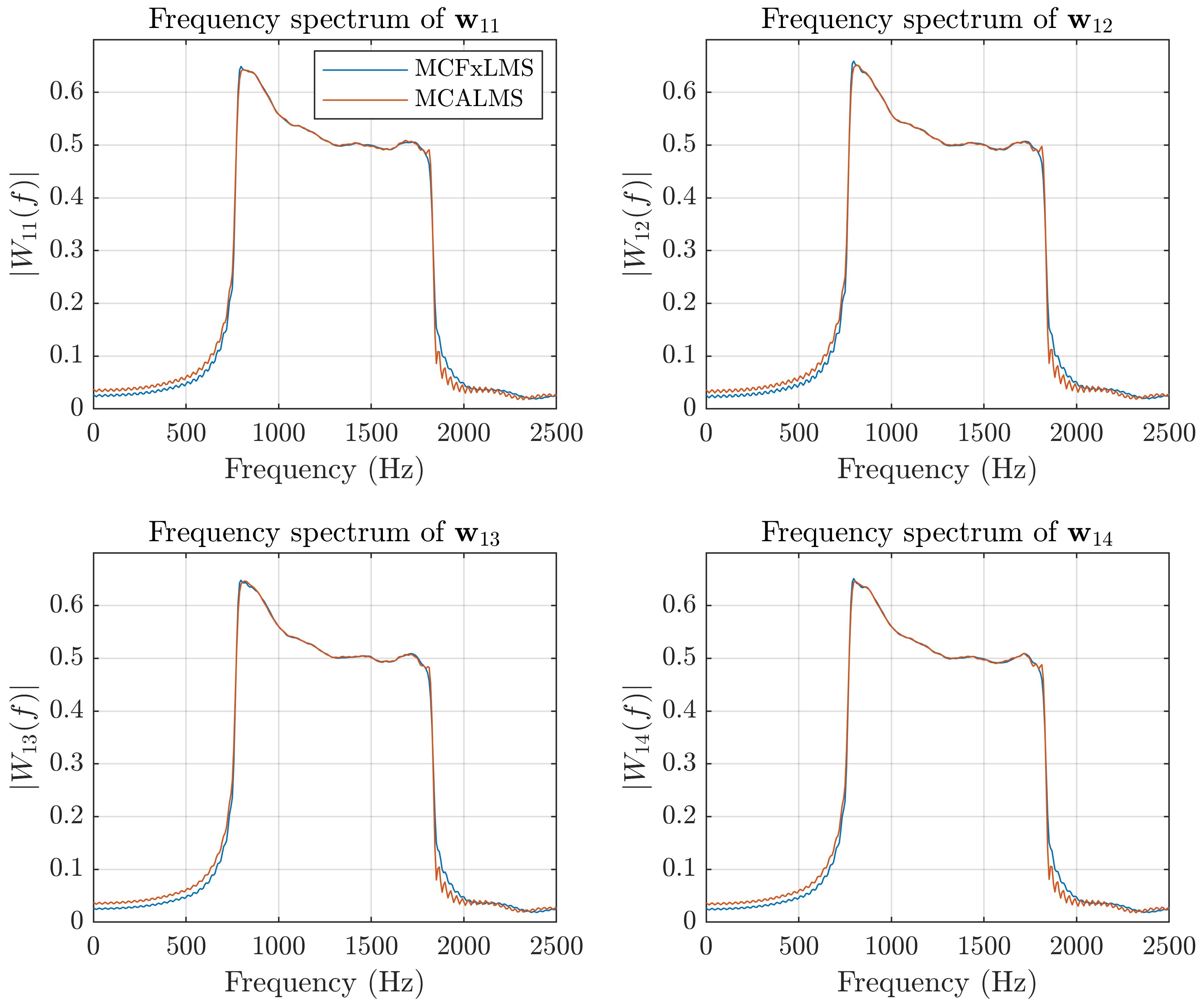}
  \end{center}
  \caption{Frequency spectrum of the control filter ${{\bm{w}}_{11}}(n)$, ${{\bm{w}}_{12}}(n)$, ${{\bm{w}}_{13}}(n)$ and ${{\bm{w}}_{14}}(n)$ in the MVANC system with MCFxLMS and MCALMS algorithms.}
  \label{fig:7}
\end{figure}

\subsection{Comparison of MVANC system utilizing MCFxLMS versus MCALMS algorithm}
\noindent
To evaluate the noise reduction capabilities of the MVANC system employing the MCALMS algorithm in comparison with the system using the MCFxLMS algorithm, broadband noises spanning a frequency range of $800-1800$ Hz are selected as the primary noise sources for both the tuning and control stages. The error signal measured at virtual microphone $1$ with these algorithms is illustrated in Fig.~\ref{fig:6}. This comparison reveals both algorithms achieve similar noise reduction efficiencies, with around $35$ dB of noise attenuation observed at steady-state. This similarity in noise reduction is attributable to the similar control filters developed by the two algorithms, as demonstrated in Fig.~\ref{fig:7}, which presents the frequency spectrum for control filters ${{\bm{w}}_{11}}(n)$, ${{\bm{w}}_{12}}(n)$, ${{\bm{w}}_{13}}(n)$ and ${{\bm{w}}_{14}}(n)$. Moreover, the remaining four control filters produced by these algorithms are almost identical. Consequently, these simulation outcomes verify that both the MCALMS and the MCFxLMS algorithms are capable of achieving equivalent optimal noise cancellation performance.

\begin{figure}[h!]
\begin{center}
  \includegraphics[width=4.5in]{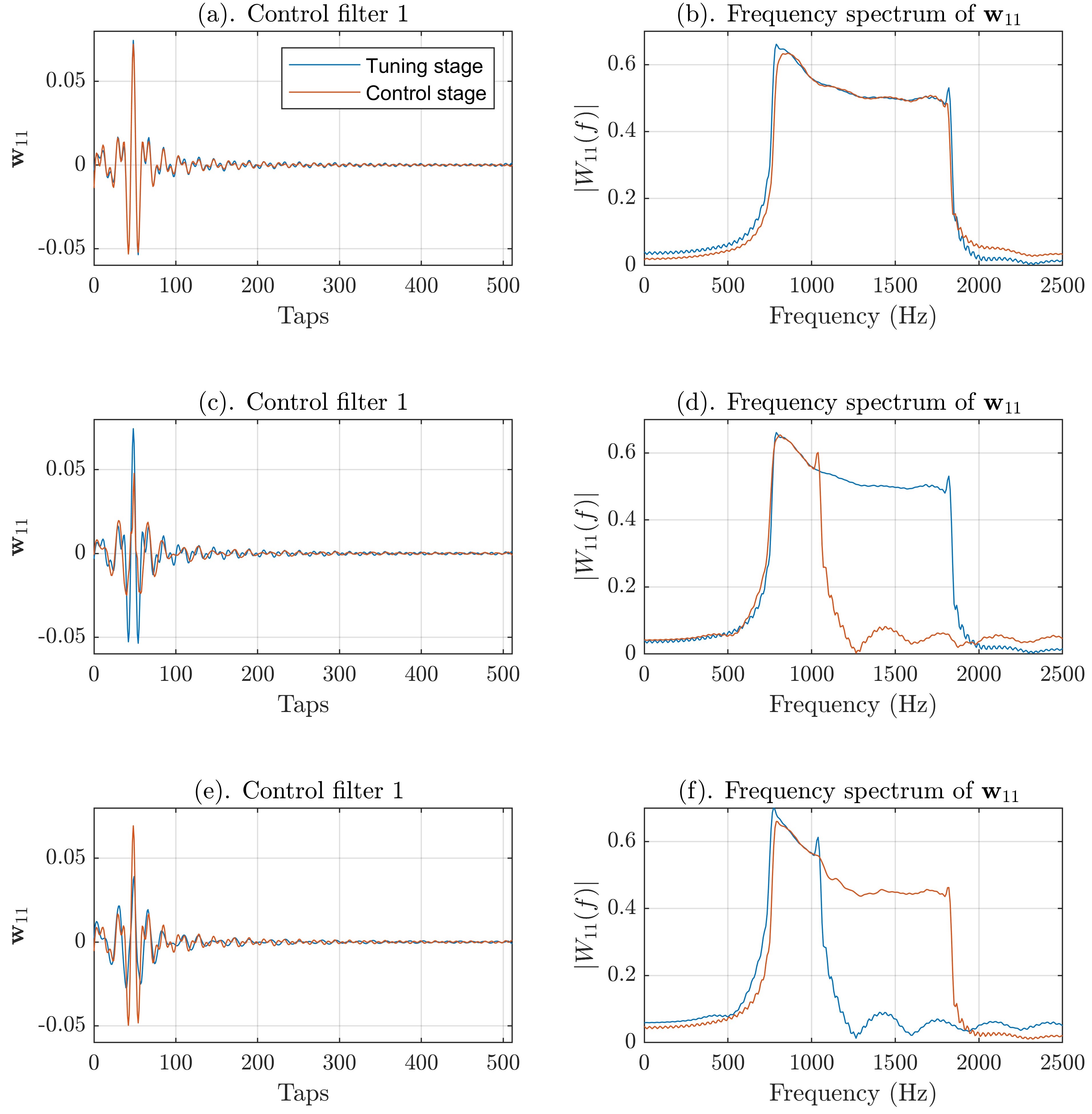}
  \end{center}
  \caption{Impulse response (left) and frequency spectrum (right) of the steady-state control filter ${{\bm{w}}_{11}}(n)$ in three scenarios: (a-b) controlling uniformly distributed noise with Gaussian tuning noise; (c-d) controlling narrowband noise with broadband tuning noise; (e-f) controlling broadband noise with narrowband tuning noise.}
  \label{fig:8}
\end{figure}

\begin{figure}[h!]
\begin{center}
  \includegraphics[width=4.8in]{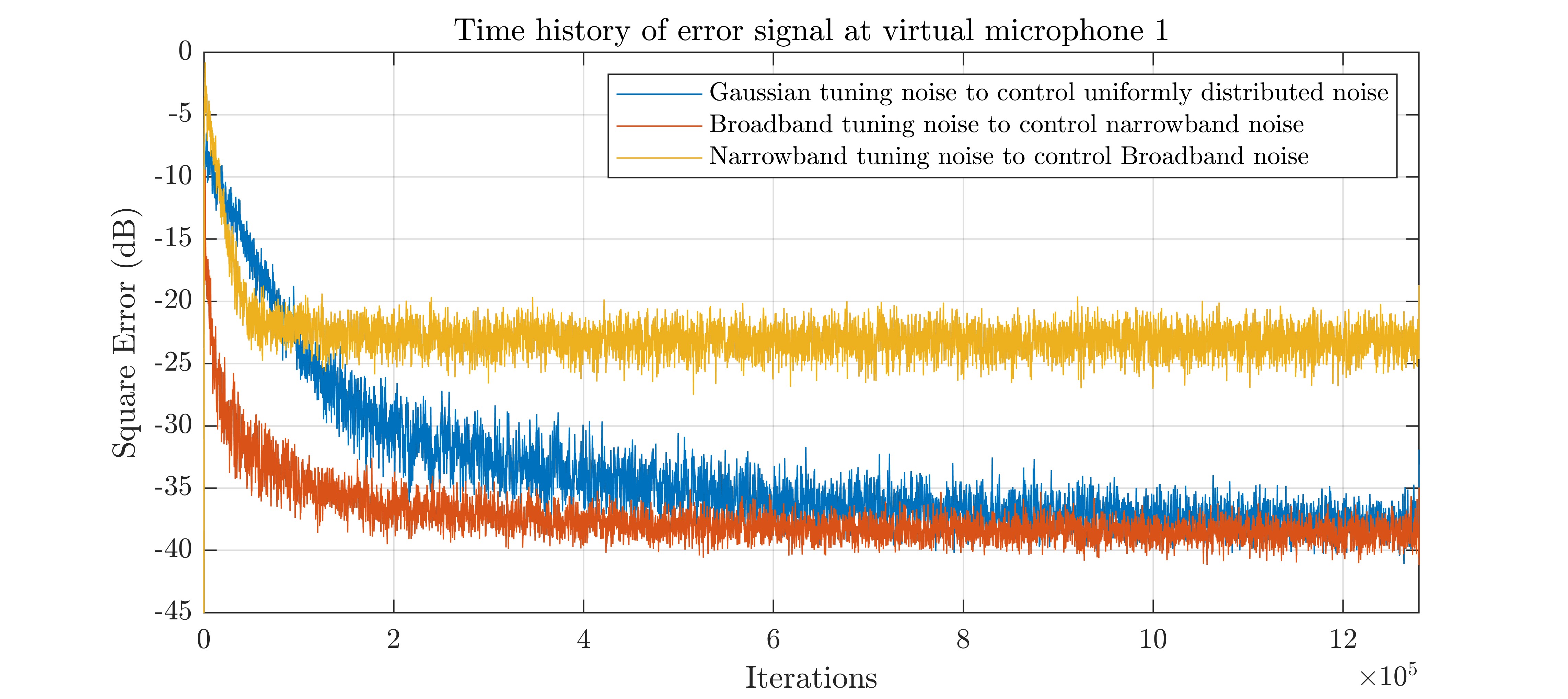}
  \end{center}
  \caption{Error signal at virtual microphone 1 of the MVANC system with MCALMS algorithm in three scenarios.}
  \label{fig:9}
\end{figure}

\subsection{Impacts of various tuning noises on noise reduction performance in the control stage}
\noindent
In the real-world application of the MVANC system, the primary noise encountered during the control stage is inherently unknown and unpredictable. This unpredictability underscores the importance of exploring how various tuning noises influence the system's noise reduction capabilities within the control stage. To this end, this simulation investigates three distinct scenarios: $(1)$ controlling uniformly distributed noise with Gaussian tuning noise, $(2)$ controlling narrowband noise with broadband tuning noise and $(3)$ controlling broadband noise with narrowband tuning noise. 

The initial scenario utilises Gaussian white noise as the tuning stage's primary noise source and uniformly distributed random noise for the control stage. Both noise types were processed through a bandpass filter spanning $800-1800$ Hz. Analysis of the control filter ${{\bm{w}}_{11}}(n)$ in both stages is depicted in Fig.~\ref{fig:8}(a) and Fig.~\ref{fig:8}(b). It can be seen that the control filter ${{\bm{w}}_{11}}(n)$ obtained from the tuning stage and the control stage are very similar and both have a passband with a frequency range of about 800-1800 Hz. Furthermore, the analysis of the other seven control filters reveals a performance analogous to that of ${{\bm{w}}_{11}}(n)$. Error signal picked up by virtual microphone $1$, shown in Fig.~\ref{fig:9}, demonstrates a steady-state noise reduction of approximately $38$ dB when using Gaussian tuning noise against uniformly distributed noise.

In the second scenario, both the tuning and control stages employ Gaussian white noise filtered through a bandpass filter; however, the frequency ranges differed: $800-1800$ Hz for tuning and $800-1000$ Hz for control. Fig.~\ref{fig:8}(c) and Fig.~\ref{fig:8}(d) showcase the control filter ${{\bm{w}}_{11}}(n)$ for each stage, highlighting the adaptation to the respective noise bandwidths. Observations reveal that control filter ${{\bm{w}}_{11}}(n)$, derived from the tuning stage, exhibits a passband within the $800-1800$ Hz frequency range. Conversely, control filter ${{\bm{w}}_{11}}(n)$ from the control stage demonstrates a passband within the narrower $800-1000$ Hz range. Despite the variance in passband ranges, the subsequent seven control filters closely resembled the control filter ${{\bm{w}}_{11}}(n)$. As Fig.~\ref{fig:9} illustrates, employing broadband tuning noise to mitigate narrowband noise resulted in a noise reduction performance of about $40$ dB.

The final scenario reversed the conditions of the second, with the tuning stage's noise filtered to $800-1000$ Hz and the control stage's to $800-1800$ Hz. Control filter ${{\bm{w}}_{11}}(n)$ for both the tuning and control stages is depicted in Fig.~\ref{fig:8}(e) and Fig.~\ref{fig:8}(f). Analysis shows that Control filter ${{\bm{w}}_{11}}(n)$ from the tuning stage encompasses a passband frequency range of approximately $800-1000$ Hz. In contrast, control filter ${{\bm{w}}_{11}}(n)$ from the control stage covers a broader range of about $800-1800$ Hz, though the passband experiences attenuation in the $1000-1800$ Hz range. Similarly, the alignment with control filter ${{\bm{w}}_{11}}(n)$ is again observed in the remaining filters. As depicted in Fig.~\ref{fig:9}, employing narrowband tuning noise to counteract broadband noise results in a steady-state noise reduction of approximately 22 dB. This diminished noise reduction performance may be due to the attenuation observed in the passband of the control filter.

This simulation confirms the effectiveness of the proposed MVANC system employing the MCALMS algorithm in achieving optimal noise suppression at virtual locations. Furthermore, it is evident that the noise reduction performance of the system is significantly enhanced when the tuning stage's primary noise frequency range encompasses that of the control stage's primary noise sources. Thus, utilizing broadband noise for auxiliary filter training is advantageous. Additionally, as demonstrated in Fig.~\ref{fig:9}, the algorithm's convergence speed improves when the noise distributions in the tuning and control stages match.

\section{Conclusion}
\noindent
The MVANC system presented in this paper leverages the MCALMS algorithm to establish an expansive quiet zone at virtual locations innovatively, addressing the practical limitations of positioning actual error microphones. Our findings reveal that with the MCALMS algorithm, the MVANC system not only achieves a significant reduction in computational complexity—up to nearly tenfold when operating with $10$ channels—but also maintains an optimal noise reduction level, equivalent to that of systems using the MCFxLMS algorithm. This demonstrates the system's ability to significantly lower computational complexity without compromising noise reduction performance. Furthermore, a series of simulations employing a variety of tuning noises have been conducted to explore the influence of the tuning noise's frequency spectrum on the noise reduction performance during the control stage. Specifically, utilizing broadband tuning noise to mitigate narrowband noise results in an impressive noise reduction level of approximately $40$ dB. Conversely, when a narrowband noise is employed to cancel broadband noise, the achieved noise reduction level diminishes to around $22$ dB. These insights serve as crucial guidelines for selecting appropriate tuning noises within the MVANC system, enhancing its noise reduction capabilities and operational efficiency.

\section*{Acknowledgements}
\noindent
This research is supported by the Singapore Ministry of Education Academic Research Fund Tier 2 (Award No.\@ MOE-T2EP50122-0018).

\bibliographystyle{unsrt}
\bibliography{sample} 

\begin{thebibliography}{10}

\bibitem{nelson1991active}
Philip~Arthur Nelson and Stephen~J Elliott.
\newblock {\em Active control of sound}.
\newblock Academic press, 1991.

\bibitem{elliott1993active}
Stephen~J Elliott and Philip~Arthur Nelson.
\newblock Active noise control.
\newblock {\em IEEE signal processing magazine}, 10(4):12--35, 1993.

\bibitem{hansen1999understanding}
Colin~H Hansen.
\newblock {\em Understanding active noise cancellation}.
\newblock CRC Press, 1999.

\bibitem{qiu2019introduction}
Xiaojun Qiu.
\newblock {\em An introduction to virtual sound barriers}.
\newblock CRC Press, 2019.

\bibitem{morgan1980analysis}
D~Morgan.
\newblock An analysis of multiple correlation cancellation loops with a filter in the auxiliary path.
\newblock {\em IEEE Transactions on Acoustics, Speech, and Signal Processing}, 28(4):454--467, 1980.

\bibitem{kuo1996active}
Sen~M Kuo and Dennis~R Morgan.
\newblock {\em Active noise control systems}, volume~4.
\newblock Wiley, New York, 1996.

\bibitem{kajikawa2012recent}
Yoshinobu Kajikawa, Woon-Seng Gan, and Sen~M Kuo.
\newblock Recent advances on active noise control: open issues and innovative applications.
\newblock {\em APSIPA Transactions on Signal and Information Processing}, 1:e3, 2012.

\bibitem{shi2024behind}
Dongyuan Shi, Woon-seng Gan, Xiaoyi Shen, Zhengding Luo, and Junwei Ji.
\newblock What is behind the meta-learning initialization of adaptive filter?—a naive method for accelerating convergence of adaptive multichannel active noise control.
\newblock {\em Neural Networks}, 172:106145, 2024.

\bibitem{ji2023practical}
Junwei Ji, Dongyuan Shi, Zhengding Luo, Xiaoyi Shen, and Woon-Seng Gan.
\newblock A practical distributed active noise control algorithm overcoming communication restrictions.
\newblock In {\em ICASSP 2023-2023 IEEE International Conference on Acoustics, Speech and Signal Processing (ICASSP)}, pages 1--5. IEEE, 2023.

\bibitem{luo2023delayless}
Zhengding Luo, Dongyuan Shi, Woon-Seng Gan, and Qirui Huang.
\newblock Delayless generative fixed-filter active noise control based on deep learning and bayesian filter.
\newblock {\em IEEE/ACM Transactions on Audio, Speech, and Language Processing}, 2023.

\bibitem{lai2023mov}
Chung~Kwan Lai, Dongyuan Shi, Bhan Lam, and Woon-Seng Gan.
\newblock Mov-modified-fxlms algorithm with variable penalty factor in a practical power output constrained active control system.
\newblock {\em IEEE Signal Processing Letters}, 2023.

\bibitem{okajima2022dual}
Ryosuke Okajima, Yoshinobu Kajikawa, and Kohei Oto.
\newblock Dual active noise control with common sensors.
\newblock In {\em ICASSP 2022-2022 IEEE International Conference on Acoustics, Speech and Signal Processing (ICASSP)}, pages 8697--8701. IEEE, 2022.

\bibitem{sun2020realistic}
Huiyuan Sun, Thushara~D Abhayapala, and Prasanga~N Samarasinghe.
\newblock A realistic multiple circular array system for active noise control over 3d space.
\newblock {\em IEEE/ACM Transactions on Audio, Speech, and Language Processing}, 28:3041--3052, 2020.

\bibitem{shen2021wireless}
Xiaoyi Shen, Dongyuan Shi, and Woon-Seng Gan.
\newblock A wireless reference active noise control headphone using coherence based selection technique.
\newblock In {\em ICASSP 2021-2021 IEEE International Conference on Acoustics, Speech and Signal Processing (ICASSP)}, pages 7983--7987. IEEE, 2021.

\bibitem{shen2022multi}
Xiaoyi Shen, Dongyuan Shi, Santi Peksi, and Woon-Seng Gan.
\newblock A multi-channel wireless active noise control headphone with coherence-based weight determination algorithm.
\newblock {\em Journal of Signal Processing Systems}, 94(8):811--819, 2022.

\bibitem{shen2022hybrid}
Xiaoyi Shen, Dongyuan Shi, and Woon-Seng Gan.
\newblock A hybrid approach to combine wireless and earcup microphones for anc headphones with error separation module.
\newblock In {\em ICASSP 2022-2022 IEEE International Conference on Acoustics, Speech and Signal Processing (ICASSP)}, pages 8702--8706. IEEE, 2022.

\bibitem{shen2022adaptive}
Xiaoyi Shen, Dongyuan Shi, Woon-Seng Gan, and Santi Peksi.
\newblock Adaptive-gain algorithm on the fixed filters applied for active noise control headphone.
\newblock {\em Mechanical Systems and Signal Processing}, 169:108641, 2022.

\bibitem{rafaely1999broadband}
B~Rafaely, SJ~Elliott, and J~Garcia-Bonito.
\newblock Broadband performance of an active headrest.
\newblock {\em The journal of the Acoustical Society of America}, 106(2):787--793, 1999.

\bibitem{chang2022multi}
Cheng-Yuan Chang, Chia-Tseng Chuang, Sen~M Kuo, and Chia-Hao Lin.
\newblock Multi-functional active noise control system on headrest of airplane seat.
\newblock {\em Mechanical Systems and Signal Processing}, 167:108552, 2022.

\bibitem{zhang2022robust}
Zeqiang Zhang, Ming Wu, Lan Yin, Chen Gong, Jun Yang, Yin Cao, and Lihua Yang.
\newblock Robust parallel virtual sensing method for feedback active noise control in a headrest.
\newblock {\em Mechanical Systems and Signal Processing}, 178:109293, 2022.

\bibitem{shen2023implementations}
Xiaoyi Shen, Dongyuan Shi, Santi Peksi, and Woon-Seng Gan.
\newblock Implementations of wireless active noise control in the headrest.
\newblock In {\em INTER-NOISE and NOISE-CON Congress and Conference Proceedings}, volume 265, pages 3445--3455. Institute of Noise Control Engineering, 2023.

\bibitem{shi2016open}
Chuang Shi, Tatsuya Murao, Dongyuan Shi, Bhan Lam, and Woon-Seng Gan.
\newblock Open loop active control of noise through open windows.
\newblock In {\em Proceedings of Meetings on Acoustics}, volume~29. AIP Publishing, 2016.

\bibitem{shi2017algorithms}
Chuang Shi, Nan Jiang, Huiyong Li, Dongyuan Shi, and Woon-Seng Gan.
\newblock On algorithms and implementations of a 4-channel active noise canceling window.
\newblock In {\em 2017 International Symposium on Intelligent Signal Processing and Communication Systems (ISPACS)}, pages 217--221. IEEE, 2017.

\bibitem{lam2018active}
Bhan Lam, Chuang Shi, Dongyuan Shi, and Woon-Seng Gan.
\newblock Active control of sound through full-sized open windows.
\newblock {\em Building and Environment}, 141:16--27, 2018.

\bibitem{lam2020active}
Bhan Lam, Dongyuan Shi, Woon-Seng Gan, Stephen~J Elliott, and Masaharu Nishimura.
\newblock Active control of broadband sound through the open aperture of a full-sized domestic window.
\newblock {\em Scientific reports}, 10(1):1--7, 2020.

\bibitem{lam2020active1}
Bhan Lam, Dongyuan Shi, Valiantsin Belyi, Shulin Wen, Woon-Seng Gan, Kelvin Li, and Irene Lee.
\newblock Active control of low-frequency noise through a single top-hung window in a full-sized room.
\newblock {\em Applied Sciences}, 10(19):6817, 2020.

\bibitem{lam2023anti}
Bhan Lam, Kelvin Chee~Quan Lim, Kenneth Ooi, Zhen-Ting Ong, Dongyuan Shi, and Woon-Seng Gan.
\newblock Anti-noise window: Subjective perception of active noise reduction and effect of informational masking.
\newblock {\em Sustainable Cities and Society}, 97:104763, 2023.

\bibitem{luo2024real}
Zhengding Luo, Dongyuan Shi, Junwei Ji, Xiaoyi Shen, and Woon-Seng Gan.
\newblock Real-time implementation and explainable ai analysis of delayless cnn-based selective fixed-filter active noise control.
\newblock {\em Mechanical Systems and Signal Processing}, 214:111364, 2024.

\bibitem{toyooka2022online}
Shota Toyooka and Yoshinobu Kajikawa.
\newblock An online identification method for virtual sensing in anc system.
\newblock {\em The Journal of the Acoustical Society of America}, 152(4):A98--A98, 2022.

\bibitem{toyooka2023hybrid}
Shota TOYOOKA and Yoshinobu KAJIKAWA.
\newblock Hybrid active noise control with auxiliary filter-based virtual sensing.
\newblock In {\em INTER-NOISE and NOISE-CON Congress and Conference Proceedings}, volume 268, pages 6488--6495. Institute of Noise Control Engineering, 2023.

\bibitem{moreau2008review}
Danielle Moreau, Ben Cazzolato, Anthony Zander, and Cornelis Petersen.
\newblock A review of virtual sensing algorithms for active noise control.
\newblock {\em Algorithms}, 1(2):69--99, 2008.

\bibitem{zhang2021robust}
Jin Zhang, Stephen~J Elliott, and Jordan Cheer.
\newblock Robust performance of virtual sensing methods for active noise control.
\newblock {\em Mechanical Systems and Signal Processing}, 152:107453, 2021.

\bibitem{liu2009virtual}
Lichuan Liu, Sen~M Kuo, and MengChu Zhou.
\newblock Virtual sensing techniques and their applications.
\newblock In {\em 2009 International Conference on Networking, Sensing and Control}, pages 31--36. IEEE, 2009.

\bibitem{garcia1997generation}
J~Garcia-Bonito, SJ~Elliott, and CC~Boucher.
\newblock Generation of zones of quiet using a virtual microphone arrangement.
\newblock {\em The journal of the Acoustical Society of America}, 101(6):3498--3516, 1997.

\bibitem{kestell2001active}
Colin~D Kestell, Ben~S Cazzolato, and Colin~H Hansen.
\newblock Active noise control in a free field with virtual sensors.
\newblock {\em The Journal of the Acoustical Society of America}, 109(1):232--243, 2001.

\bibitem{moreau2009active}
Danielle~J Moreau, Justin Ghan, BS~Cazzolato, and AC~Zander.
\newblock Active noise control in a pure tone diffuse sound field using virtual sensing.
\newblock {\em The Journal of the acoustical Society of america}, 125(6):3742--3755, 2009.

\bibitem{jung2017local}
Woomin Jung, Stephen Elliott, and Jordan Cheer.
\newblock Local active sound control using the remote microphone technique and head-tracking for tonal and broadband noise sources.
\newblock 2017.

\bibitem{jung2019local}
Woomin Jung, Stephen~J Elliott, and Jordan Cheer.
\newblock Local active control of road noise inside a vehicle.
\newblock {\em Mechanical Systems and Signal Processing}, 121:144--157, 2019.

\bibitem{hasegawa2018window}
Rina Hasegawa, Dongyuan Shi, Yoshinobu Kajikawa, and Woon-Seng Gan.
\newblock Window active noise control system with virtual sensing technique.
\newblock In {\em INTER-NOISE and NOISE-CON Congress and Conference Proceedings}, volume 258, pages 6004--6012. Institute of Noise Control Engineering, 2018.

\bibitem{shi2019analysis}
Dongyuan Shi, Bhan Lam, and Woon-seng Gan.
\newblock Analysis of multichannel virtual sensing active noise control to overcome spatial correlation and causality constraints.
\newblock In {\em ICASSP 2019-2019 IEEE International Conference on Acoustics, Speech and Signal Processing (ICASSP)}, pages 8499--8503. IEEE, 2019.

\bibitem{shi2020feedforward}
Dongyuan Shi, Woon-Seng Gan, Bhan Lam, Rina Hasegawa, and Yoshinobu Kajikawa.
\newblock Feedforward multichannel virtual-sensing active control of noise through an aperture: Analysis on causality and sensor-actuator constraints.
\newblock {\em The Journal of the Acoustical Society of America}, 147(1):32--48, 2020.

\bibitem{pawelczyk2004adaptive}
Marek Pawelczyk.
\newblock Adaptive noise control algorithms for active headrest system.
\newblock {\em Control engineering practice}, 12(9):1101--1112, 2004.

\bibitem{hasegawa2017headrest}
Rina Hasegawa, Yoshinobu Kajikawa, Cheng-Yuan Chang, and Sen~M Kuo.
\newblock Headrest application of multi-channel feedback active noise control with virtual sensing technique.
\newblock In {\em INTER-NOISE and NOISE-CON Congress and Conference Proceedings}, volume 255, pages 3513--3524. Institute of Noise Control Engineering, 2017.

\bibitem{hirose2017effectiveness}
Shun Hirose and Yoshinobu Kajikawa.
\newblock Effectiveness of headrest anc system with virtual sensing technique for factory noise.
\newblock In {\em 2017 Asia-Pacific Signal and Information Processing Association Annual Summit and Conference (APSIPA ASC)}, pages 464--468. IEEE, 2017.

\bibitem{miyazaki2015head}
Nobuhiro Miyazaki and Yoshinobu Kajikawa.
\newblock Head-mounted active noise control system with virtual sensing technique.
\newblock {\em Journal of Sound and Vibration}, 339:65--83, 2015.

\bibitem{deng2018active}
He-Siang Deng, Cheng-Yuan Chang, and Sen~M Kuo.
\newblock Active noise control with virtual sensing technology for headrest.
\newblock In {\em 2018 Asia-Pacific Signal and Information Processing Association Annual Summit and Conference (APSIPA ASC)}, pages 1272--1275. Ieee, 2018.

\bibitem{wan1996adjoint}
Eric~A Wan.
\newblock Adjoint lms: An efficient alternative to the filtered-x lms and multiple error lms algorithms.
\newblock In {\em 1996 IEEE International Conference on Acoustics, Speech, and Signal Processing Conference Proceedings}, volume~3, pages 1842--1845. IEEE, 1996.

\bibitem{shi2023computation}
Dongyuan Shi, Bhan Lam, Junwei Ji, Xiaoyi Shen, Chung~Kwan Lai, and Woon-Seng Gan.
\newblock Computation-efficient solution for fully-connected active noise control window: Analysis and implementation of multichannel adjoint least mean square algorithm.
\newblock {\em Mechanical Systems and Signal Processing}, 199:110444, 2023.

\end{thebibliography}

\end{document}